\documentclass[11pt]{article}
\usepackage{cite}
\usepackage{amsmath,amsfonts,amssymb}
\usepackage[small,bf,hang]{caption}
\usepackage{epsfig}
\usepackage{slashed}
\usepackage{mathabx,amsmath}
\usepackage{color}
\usepackage{verbatim}

\def\hybrid{
        \topmargin -20pt
        \oddsidemargin 0pt
        \headheight 0pt \headsep 0pt
        \textwidth 6.25in 
        \textheight 9.5in 
        \marginparwidth .875in
        \parskip 5pt plus 1pt \jot = 1.5ex}

\hybrid

 \csname
@addtoreset\endcsname{equation}{section}

\newcommand{\Exp}[1]{\exp\left(#1\right)}

\newcommand{\mbar}{\overline{\cM}}

\newcommand{\fd}[2]{\parbox{#1}{\includegraphics[width=#1]{#2}}}

\newcommand{\R}{\mathbb{R}}
\newcommand{\CP}{\mathbb{CP}}

\newcommand{\abs}[1]{\left|#1\right|}

\def\moth{\mathsurround=0pt}
\newdimen\zo \zo=0pt

\def\tick{\leaders\hrule height 0.5ex depth 0pt \hskip 0.5pt}
\def\upboxfill{$\moth \setbox\zo\hbox{\tick}%
  \hskip 3pt\hbox to 0pt{$\tick$\hss}\hrulefill \hbox to 7.5pt{$\tick$\hss}$}

\def\dtick{\leaders\hrule height .34pt depth 0.5ex \hskip 0.5pt}
\def\downboxfill{$\moth \setbox\zo\hbox{\dtick}%
  \hskip 2pt\hbox to 0pt{$\dtick$\hss}\hrulefill \hbox to 2pt{$\dtick$\hss}$}

\def\bec{\begin{center}}
\def\ec{\end{center}}

\def\C{\Gamma}

\def\cF{{\cal F}}
\def\cO{{\cal O}}
\def\chP{\widehat{\cal P}}

\def\cM{{\cal M}}
\def\cP{{\cal P}}

\def\cV{{\cal V}}

\def\cW{{\cal W}}

\def\cO{{\cal O}}

\renewcommand{\i}{\mathsf{i}}
 
\def\be{\begin{equation}}
\def\ee{\end{equation}}
\def\bea{\begin{eqnarray}}
\def\eea{\end{eqnarray}}
\def\ba{\begin{array}}
\def\ea{\end{array}}

\begin{document}

\begin{titlepage}
\rightline{}
\rightline{August 2019}
\rightline{MIT-CTP-5139}       
\begin{center}
\vskip 1.5cm
 {\Large \bf{ Hyperbolic String Vertices }}
\vskip 1.7cm

{\large\bf {Kevin Costello${}^{1}$ and Barton Zwiebach${}^2$}}
\vskip 1.6cm

{\it ${}^1$ Perimeter Institute of Theoretical Physics,\\
 31 Caroline St N, Waterloo, ON N2L 2Y5, Canada}\\
\vskip .1cm

\vskip .2cm

{\em $^2$ \hskip -.1truecm Center for Theoretical Physics, \\
Massachusetts Institute of Technology\\
Cambridge, MA 02139, USA \vskip 5pt }

\end{center}

\bigskip\bigskip
\begin{center} 
\textbf{Abstract}

\end{center} 
\begin{quote}  
\small The string vertices of closed string field theory
are subsets of the moduli spaces of punctured
Riemann surfaces that satisfy a geometric version of the 
Batalin-Vilkovisky master equation.   
We present a homological proof of existence of string vertices
and their uniqueness up to canonical transformations.
Using hyperbolic metrics on surfaces  
with geodesic boundaries we give  
an exact construction of string vertices as sets of  surfaces  
with systole greater than or equal   
to~$L$ with $L\leq 2\,  \hbox{arcsinh} \,1$.  
Intrinsic hyperbolic collars prevent the 
appearance of short geodesics upon sewing.  
The surfaces generated by Feynman diagrams are naturally endowed with 
Thurston metrics: hyperbolic on the vertices
and flat on the propagators.  For the classical theory the 
length $L$  is arbitrary and, as $L\to \infty$  
 hyperbolic vertices become the minimal-area vertices
of closed string theory.  

\end{quote} 
\vfill
\setcounter{footnote}{0}
\end{titlepage}

\tableofcontents
\baselineskip 14pt 


\section{Introduction}

The key geometric input for the construction of string field theories
is a set of string vertices.  For the case of closed string field theories,
including heterotic and type II strings, 
string vertices $\cV_{g,n}$ are subsets of the moduli spaces of compact
Riemann surfaces of genus $g$ and $n$ marked points, with a choice of local
coordinates (defined up to phases) at those marked points.  At genus zero
string vertices are required for $n\geq 3$, at genus one for $n\geq 1$, and
for genus two or greater, for $n \geq 0$. 
If consistent string vertices are known,  a choice of a suitable conformal 
field theory allows the construction of bosonic closed string field theory\cite{Zwiebach:1992ie}.   With a choice of 
suitable superconformal field theories, a proper set of string fields, and 
careful distributions of picture changing operators,  it is now
known how to  use string vertices to construct all closed superstring 
theories~\cite{Sen:2015uaa,deLacroix:2017lif}. 
  
String vertices lead to closed string field theories that satisfy the
Batalin-Vilkoviski (BV) master equation, and are therefore consistent quantum
theories,
if they satisfy a  geometric version of the master 
equation~\cite{Sen:1994kx,Sen:1993kb,Sonoda:1989wa}.  This {\em geometric
master equation} reads:
\be
\label{m-eqn}
\partial \cV + \hbar \, \Delta \cV  + \tfrac{1}{2} \{ \cV\,, \cV \} = 0 \,, 
\ee
for sets that comprise $\cV$ as follows
\be
\label{formal-sum}
\cV = \sum_{g,n} \hbar^g\,  \cV_{g,n} \,, \quad \hbox{with}
\quad 
\begin{cases}  n \geq 3\,, \ \hbox{for} \ \, g = 0 \,, \\
n\geq 1\,, \ \hbox{for} \ \, g= 1 \,, \\
n\geq 0 \,, \ \hbox{for} \ \,  g \geq 2\,. 
\end{cases}
\ee
Interestingly, the list of vertices above is precisely that for which
the surfaces have negative Euler number and thus admit hyperbolic
metrics of constant negative curvature. 
The various operations in the master equation were defined 
in~\cite{Sen:1994kx,Sen:1993kb}. Briefly, $\partial$ denotes boundary, $\Delta$ involves removing 
coordinate disks about two marked points on a Riemann surface, and
then sewing and twisting the boundaries.  Finally, the two-input `anti bracket' $\{ \cdot , \cdot \}$  
takes a surface from each input, 
removes a coordinate disk from each, sews the boundaries and twists.  
Twist sewing of two
local coordinates $z_1, z_2$ means removing the $|z_1| < 1, |z_2|< 1$ 
disks and gluing the $|z_1| = 1$ and $|z_2|=1$ boundaries via $z_1 z_2 = e^{i\theta}$ for
all $\theta \in [0, 2\pi)$. 
The $\cV_{g,n}$ are subsets of the bundle $\chP_{g,n}$ 
over the moduli space $\cM_{g,n}$ of Riemann surfaces of genus
$g$ and $n$ marked points. 
A point in $\chP_{g,n}$ is a Riemann  surface $\Sigma_{g,n}$ 
of genus $g$ and with $n$ marked points, with local coordinates defined
up to a phase around the points.  The local coordinates are 
represented as embedded disks on $\Sigma_{g,n}$
that do not overlap. 
 The bundle $\chP_{g,n}$ comes with a natural projection $\pi$ 
to $\cM_{g,n}$:
\be
\pi : \ \chP_{g,n}\to \cM_{g,n}\,.
\ee
The map $\pi$ forgets the local coordinates at the punctures.  
String vertices were used to construct the partition function of certain topological string theories~\cite{Costello}.  

\begin{figure}[!ht]
	\centering
	\fd{11.0cm}{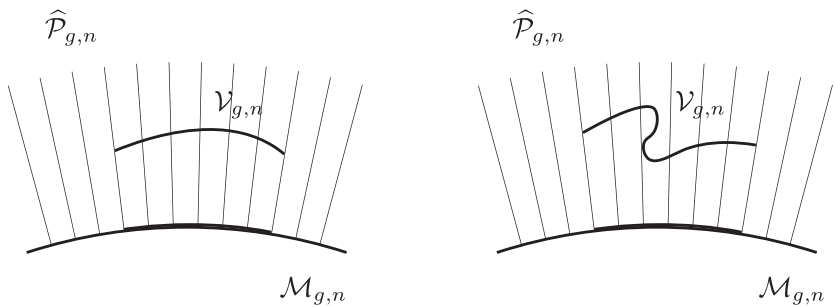}  
	\caption{Left:  A string vertex $\cV_{g,n}$ that is a piece of a section
	over $\cM_{g,n}$.  Right:  A string vertex $\cV_{g,n}$ that is not a piece
	of a section over $\cM_{g,n}$.   
	In both cases the projection $\pi$ maps $\cV_{g,n}$ to its image with
	degree one.}
	\label{fig:sect-y-no}
\end{figure}

The $\cV_{g,n}$ are traditionally required to be  
pieces of {\em sections} on the bundle $\chP_{g,n}$ 
over $\cM_{g,n}$.   This condition, if satisfied, affords
a degree of simplicity, and makes the mathematical construction
more canonical, as each underlying Riemann surface in the vertex
is constructed once and only once.  It seems clear, however, 
that the consistency of the string field theory is satisfied by vertices 
that are more general and are not strict sections. 
All that is needed is that $\cV_{g,n}$ maps 
onto its image under $\pi$ 
with degree one. This means that a generic surface 
is counted once \emph{with multiplicity}. 
The situation is sketched in 
Figure~\ref{fig:sect-y-no}.    

 When the string vertices are used together with a propagator to form
Feynman graphs of string field theory, the result is some 
{\em submanifold}  
$\cF_{g,n}$ in $\chP_{g,n}$.  It is traditionally required 
that $\cF_{g,n}$ should be a full section of the bundle $\chP_{g,n}$, 
so that the map $\cF_{g,n} \to \cM_{g,n}$ is a homeomorphism.  
Again, this is not necessary: all that is required is that the map 
$\cF_{g,n} \to \cM_{g,n}$ is of degree one. Such a map is surjective.   

This is enough to imply that the integral over $\cF_{g,n}$ of any differential form pulled back from $\cM_{g,n}$ will be the same as the integral over $\cM_{g,n}$.  
Since the integrand for on-shell string states is a top-form on $\cM_{g,n}$, this implies that the amplitude for on-shell string states computed by integrating over $\cM_{g,n}$ coincides with that computed using string field theory Feynman rules.  

It is sometimes useful to consider string vertices which are not sub-manifolds of $\chP_{g,n}$, but rather singular chains of degree $6g-6+2n$.  This is reasonable, because the purpose of introducing string vertices is to integrate over them, and there is a well-behaved theory of integration pairing differential forms with singular chains.  In this generality, the string vertices $\cV_{g,n}$ must satisfy two constraints:
\begin{enumerate}

\item The chain $\cV$ must satisfy the geometric master equation~\eqref{m-eqn}. 

\item 
The chain $\cF_{g,n}$, constructed as Feynman diagrams using $\cV$ and a propagator, must, when pushed forward to $\cM_{g,n}$, represent the fundamental homology class\footnote{In the homology relative to the boundary of $\cM_{g,n}$, or equivalently in the homology of the Deligne-Mumford compactification.} of $\cM_{g,n}$.  

\end{enumerate}
In the case that $\cF_{g,n}$ is represented by a submanifold of $\chP_{g,n}$, this second constraint implies that the map $\cF_{g,n} \to \cM_{g,n}$ is of degree one. Without the assumption that $\cF_{g,n}$ is a submanifold, this constraint implies that the integral over $\cF_{g,n}$ of a differential form on $\chP_{g,n}$ which is pulled  back from $\cM_{g,n}$ coincides with the integral over $\cM_{g,n}$.

For string vertices in this sense, existence and uniqueness (up to canonical transformations) was proved in~\cite{Costello}.  
We give a largely self-contained review and detailed 
elaboration of these arguments 
in section~\ref{sec:homological}.     
The earlier discussion of these matters in the physics literature~\cite{Hata:1993gf,Sen:1994kx} did not address a priori existence,  and uniqueness was shown assuming 
the vertices are submanifolds and partial sections of $\chP_{g,n}$.   
We also show that the second condition listed above holds for
any string vertices that satisfy the first condition.  
Thus satisfying the master equation is {\em all} that is required of $\cV$.  

One approach to determine the string vertices explicitly is through a (conformal) minimal
area problem~\cite{Zwiebach:1990nh}.  On each Riemann
surface in $\cM_{g,n}$,  the problem asks for the metric
of least area for a fixed systole (the length of the
shortest closed geodesic).  The minimal area metrics for genus zero
and $n \geq 3$ are well known and arise from Jenkins-Strebel quadratic
differentials.  For genus one or greater, the minimal area metrics
are partially known.  As of now there is no general proof of existence
of the minimal area metrics, presumably because it is not known how singular
the metrics can be.  Recently, however, Headrick and one of us used convex optimization to develop theoretical and computational
tools to deal with these minimal area metrics, to find accurate numerical
solutions for previously unknown metrics~\cite{Headrick:2018ncs,Headrick:2018dlw}
and, with Naseer, 
 to obtain insight into the closely related Riemannian isosytolic problem~\cite{Naseer:2019zau}.  The new metrics do not seem to be particularly
singular, and one could expect that an existence proof will be developed in
the near future.   Using the minimal area metrics 
there is a simple prescription to define string vertices that satisfy the
geometric master equation.  Since minimal area metrics are unique, the vertices
are pieces of sections over $\cM_{g,n}$ and the Feynman rules generate
full sections over~$\cM_{g,n}$.    

A new and intriguing approach  
to the question of finding string 
vertices  was recently developed by Moosavian and Pius~\cite{Moosavian:2017qsp,Moosavian:2017sev} using the framework of hyperbolic geometry.  
The use of hyperbolic geometry is particularly attractive due to recent
progress using  Teichm\"uller spaces of hyperbolic metrics 
in order to compute integrals over the moduli spaces of Riemann 
surfaces~\cite{mc-shane,mirz1,mirz2} (for a review of some  
of these ideas see~\cite{Dijkgraaf:2018vnm}).  The authors considered hyperbolic 
metrics (metrics of Gaussian curvature $K= -1$) 
on {\em punctured} Riemann surfaces in $\cM_{g,n}$ and defined subsets
that to first approximation are string vertices.  The metrics
have cusps at the punctures and can be used to define 
local coordinates around them.  The boundaries of the coordinate disks, however, are horocycles rather than geodesics.   This implies that the sewing of these 
surfaces do not yield exactly hyperbolic metrics.  The vertices and local coordinates
around the punctures must then be corrected, and the authors discuss how
to do this to first order in a cutoff parameter using results by Wolpert~\cite{wolpert}
and Obitsu and Wolpert~\cite{wolpert2}.

Motivated by their work and following a suggestion of one of us~\cite{k-costello},   
we give here a simple and explicit hyperbolic construction of string vertices. 
 Instead of starting with $\cM_{g,n}$, however, we use 
 moduli spaces of hyperbolic metrics on surfaces of genus $g$ and $n$ 
 {\em geodesic boundaries}.   
 By restricting the metrics to those for which the boundaries all have
 length $L$ we consider the moduli space $\cM_{g,n, L}$.  
On these surfaces the systole is defined as the shortest geodesic  that  
 is {\em not} a boundary component. 
We define string vertices as {\em systolic} 
 subsets of these moduli spaces: they include all surfaces
 with systole greater than or equal to $L$.  Namely, surfaces belonging to a vertex
 have no geodesics of length less than $L$. 
For each surface
 in the vertices we turn the $n$ boundaries into $n$ punctured
 disks by  grafting flat semi-infinite cylinders
 of circumference~$L$.  This is a canonical operation giving us
 a homeomorphism from $\cM_{g,n, L}$ to $\cM_{g,n}$. 
It is not hard to show that for any $L < L_* =  2\, \hbox{arcsinh}  \, 1$ 
the string vertices satisfy {\em exactly} the geometric master equation. 
As a result the vertices define a closed string field theory that satisfies
the BV master equation exactly.  
A key role in the construction 
is played by the collar theorems of hyperbolic geometry~\cite{buser}: the collars about the geodesic boundaries play the same role as stubs do in minimal
area metrics: they prevent the creation closed curves shorter than $L$ upon sewing. 
Vertices with different values of $L$ 
are equivalent up to BV canonical transformations.
 
We use the work of Mondello~\cite{mondello}  to explain that the hyperbolic string vertices are in fact pieces of sections in $\chP_{g,n}$.
To form the Feynman diagrams of this theory, one joins the boundaries
on surfaces belonging to string vertices by attaching 
annuli, representing the propagator.  The string field theory does not prescribe
a metric on the annuli, but it is clear that making the
annulus into a flat cylinder of circumference~$L$ defines a continuous metric
throughout the surface.  Note that a hyperbolic metric on the annulus would
give a metric that is discontinuous at the seams and is therefore
 not hyperbolic on the whole surface.  
The mixed metric, hyperbolic on the vertices and flat on the cylinders, is natural.  It arises in the theory
of $\mathbb{CP}^1$ projective structures on surfaces, and in the 
definition of grafting across measured geodesic laminations.  In such
theory the above metric, called the {\em Thurston metric}, arises naturally
as the Kobayashi metric in the category of $\mathbb{CP}^1$ surfaces.
Results by Dumas and Wolf~\cite{dumas_wolf}  hint that
we may be also getting sections from the Feynman region.  

\noindent
 {\em Note added:}  We were informed by J\o rgen E. Andersen that 
he also showed that the systolic sets suggested in~\cite{k-costello}
solve the quantum master equation~\cite{j-ellegard}.
  
\section{ Existence and uniqueness of string vertices}
\label{sec:homological} 

In this section we will review the homological argument of~\cite{Costello} that proves the existence of the string vertices, and their uniqueness up to canonical transformation. (The super-string version of this argument has recently been provided by Moosavian and Zhou~\cite{moosavian_zhou}).   
We also show how the construction guarantees that 
the chain constructed as Feynman diagrams represents the fundamental homology class of $\cM_{g,n}$.

The material in this section provides perspective on the problem of finding
explicit string vertices, but is not required for the
 construction of hyperbolic
vertices in the subsequent sections.

\subsection{Canonical transformations}
Suppose that we have some collection $\cV_{g,n}$ of string vertices. For this section, we do not require that the $\cV_{g,n}$ are represented by submanifolds of $\chP_{g,n}$. All we ask is that the $\cV_{g,n}$ are singular chains with real coefficients, that is elements of $C_{6g-6+2n}(\chP_{g,n})$.  

If we have some collection $\cW_{g,n} \in C_{6g-6+2n+1}(\chP_{g,n})$, then we can vary the string vertices $\cV_{g,n}$~by
\begin{equation} 
	\delta_{\cW} \cV = 
	\{ \cV\,, \cW \}  
	+ \hbar \Delta \cW + \partial \cW \,, 
\end{equation}
where, as before, $\cV = \sum \hbar^g \cV_{g,n}$, $\cW = \sum \hbar^g \cW_{g,n}$.  It is automatic that this variation of $\cV$ still satisfies the quantum master equation, to leading order in $\delta_{\cW} \cV$:\footnote{Useful identities $\Delta^2 = \partial^2 = 0$,   $\{ X, Y\} =  -(-1)^{(X+1)(Y+1)} \{ Y , X\}$, $\  \Delta \partial X = - \partial \Delta X$,  $\ \partial\{ X, Y\} 
= \{ \partial X, Y\} + (-1)^{X+1} \{ X, \partial Y \}$,  $\ \Delta\{ X, Y\} 
= \{ \Delta X, Y\} + (-1)^{X+1} \{ X, \Delta Y \}$ and $(-1)^{(X_1 +1)(X_3+1)} 
\bigl\{ \{ X_1, X_2\} , X_3 \bigr\}  + \hbox{cyclic} = 0$.  The values in exponents
are degrees, given by the dimension of the space.  All $\partial, \Delta$, and $\{ \cdot, \cdot \} $ change degree by one unit.}
\begin{equation} 
\begin{split}
	 & \tfrac{1}{2}  
	\{\cV + \delta_{\cW} \cV, \cV + \delta_{\cW} \cV\} 
	+ \hbar  \Delta  (\cV + \delta_{\cW} \cV) + \partial (V + \delta_{\cW} \cV) \\
	  &= \bigl\{ 
	  \partial \cV + \hbar \, \Delta \cV  + \tfrac{1}{2} \{ \cV\,, \cV \} 
	  \,,  \cW   \bigr\}  + O( (\delta_{\cW} \cV)^2 )  =  0  \,.
\end{split}
\end{equation}
A variation of this form is an infinitesimal canonical transformation.

Two sets of string vertices which are related by canonical transformations like this yield equivalent string field theories.  To understand this, let us recall some background on how the string field theory action is constructed.

The string action $S_\cV$ associated to a set of vertices is given by
\be
\label{action-bv}
S_\cV (\psi)   =  S_{0,2}  + I_\cV (\psi)\,,
\ee
where $S_{0,2} = {1\over 2} \langle \psi, Q\psi\rangle$ is the kinetic term and 
does not depend on the vertices.  
Moreover, $I_\cV (\psi)$ denotes the interactions, obtained by integrating 
the correlators of the closed string field $\psi$ over the moduli spaces in $\cV$.
The compatibility of the BV anti-bracket in string field theory with that defined by cutting and gluing Riemann surfaces tells us that
for chains $A, B$ we have~\cite{Sen:1993kb}:
\be
I_{\Delta A }  = - \Delta I_A \,,   \quad I_{\{ A, B\}} = - \{  I_A, I_B\} \,, \quad
I_{\partial A} = - \{ S_{0,2},  I_A\} \,. 
\ee
It now follows from~(\ref{action-bv}) and the expression for $\delta_\cW \cV$
that 
\be
\begin{split}
S_{\cV + \delta_\cW \cV} = & \   S_\cV  +  I_{\delta_\cW \cV} \\
= & \   S_\cV  +  I_{\{ \cV\,, \cW \} + \hbar \Delta \cW + \partial \cW   } \\
= & \   S_\cV - \{  I_\cV\,, I_\cW \} -  \hbar \Delta I_\cW - \{ S_{0,2},  \cW  \} \\ 
= & \   S_\cV - \{  S_\cV\,, I_\cW \} -  \hbar \Delta I_\cW \,,
\end{split}
\ee
showing that the action for the new vertices
is obtained by an infinitesimal
canonical transformation of the original action.  Indeed, the transformation
$\delta S =  \hbar \Delta \epsilon + \{ S, \epsilon\}$ with $\epsilon$
an odd parameter leaves the master equation
$\hbar \Delta S+ 
{1\over 2} \{  S, S\} =0$ unchanged.   This shows that vertices related
by a canonical transformation can be regarded as equivalent. 
Observables transform under canonical transformations and 
their expectation values are also unchanged.

\subsection{Existence of string vertices}
The argument for existence of string vertices is inductive. We start by fixing $\cV_{0,3}$, which is of dimension $0$, to be $\mathbb{CP}^1$ with marked points $0,1,\infty$ and some choice of coordinates around the marked points,
invariant under the action of $S_3$ permuting the marked points.  

For the inductive step, fix a pair of integers $(g,n)$. We assume, by induction, that we have constructed $\cV_{g',n'}$ for all $(g',n')$ with $g' < g$, or with $g' = g$ and $n' < n$. We assume that the $\cV_{g',n'}$ we have constructed satisfy the master equation.  We also assume that $\cV_{g',n'}$ is invariant under the action of $S_{n'}$ permuting the marked points.

We need to find $\cV_{g,n}$ satisfying
\begin{equation} 
	\partial \cV_{g,n} = - \Delta \cV_{g-1,n+2}\  -\ 
	 \tfrac{1}{2} \hskip-10pt\sum_{\substack{g_1 + g_2 = g \\ n_1 + n_2 = n+2} } 
	 \hskip-5pt\{\cV_{g_1,n_1}, \cV_{g_2,n_2} \} 
	 \ \equiv  \cO_{g,n} \,.   
	  \label{eqn:qme} 
\end{equation}
The right hand side $\cO_{g,n}$ of this equation is a 
$(6g-6+2n)-1$    chain on $\chP_{g,n}$.  It is build entirely from $\cV_{g',n'}$ which we have already constructed in our induction.   By construction, $\cO_{g,n}$ is an $S_n$ invariant chain, as required for $\cV_{g,n}$ to be $S_n$ invariant as well.

Our inductive assumption that the $\cV_{g',n'}$ satisfy the quantum master equation implies that
\begin{equation} 
	 \partial \cO_{g,n} = 0 \,.   
\end{equation}
This can be  checked by writing $\partial \cV = - \hbar \Delta \cV - {1\over 2} \{ 
\cV, \cV \} $ and confirming that $\partial$ kills the right-hand side using 
 $\{ \{ \cV, \cV\} , \cV\} =0$.  
The problem of constructing the string vertex $\cV_{g,n}$ satisfying \eqref{eqn:qme}  amounts to showing that the homology class 
\begin{equation} 
	[\cO_{g,n}] \in H_{6g-6+2n-1}(\chP_{g,n})^{S_n} 
\end{equation}
vanishes.  The superscript $S_n$ indicates that we are focusing on the homology
of $S_n$-invariant chains.\footnote{If the   
unrestricted homology group vanishes,
the homology of $S_n$-invariant chains will vanish as well. But if the unrestricted
homology group does not vanish, it may still vanish for $S_n$-invariant chains.} 
 
To prove that $[\cO_{g,n}] = 0$, we will show that the homology group $H_{6g-6+2n+1}(\chP_{g,n})^{S_n}$ is zero.  The proof of this consists of two steps.
\begin{enumerate} 
	\item First, we show that $\chP_{g,n}$ is homotopy equivalent to $\cM_{g,n}$.  This implies that the homology groups of $\chP_{g,n}$ and of $\cM_{g,n}$ are isomorphic. 
	\item Then, we show that $H_{6g-6+2n-1}(\cM_{g,n})^{S_n}$ is zero. 
\end{enumerate}

\subsubsection{Proving the homotopy equivalence $\chP_{g,n} \simeq \cM_{g,n}$}
\label{pro-the-hom-equ}


Let us recall that $\chP_{g,n}$ is the moduli space of Riemann surfaces $\Sigma$ with $n$ embedded discs $D_1,\dots,D_n$.  We assume here that the closures of the $D_i$ are disjoint. On each disc, we have a coordinate function $z_i$, defined up to rotation, such that $D_i = \{ |z_i| \leq 1\}$, 
and we assume that~$z_i$  
extends analytically to a small neighbourhood in $\Sigma$ of the closure $\overline{D}_i$ of the disc.   

This moduli space can be described equivalently as follows.  Consider the moduli space $\chP'_{g,n}$ of Riemann surfaces with $n$ boundary components $b_i$, and a coordinate $\theta_i$ on each boundary component, defined up to a shift. We assume that the coordinate $\theta_i$ has the following properties:
\begin{enumerate}
	\item $\theta_i$ has range $2 \pi$.
	\item $\theta_i$ is real analytic.
	\item The vector field $\partial_{\theta_i}$ points in the direction given by the orientation of the boundary component $b_i$ which is induced from the orientation of the bulk surface. 
\end{enumerate}

Suppose, near some point in one of the boundary components, we choose coordinates $z = x+ \i y$ where the boundary is at $y = 0$ and a local patch of the surface is in the upper half-plane $y \ge 0$. Then, $\theta_i$ is some function $F_i(x)$. 
Our assumption that $\theta_i$ is real-analytic means $F_i$ is a real-analytic function of $x$, and so represented by a convergent power series.  Because of this, 
$F_i(z)$ will also converge in some domain.  This implies that $\theta_i$ is the boundary value of the holomorphic function $F_i(z)$, and because $\theta_i$ provides a coordinate on the boundary, $F_i(z)$ must provide a holomorphic coordinate in some neighbourhood of the boundary.   This fact will be key for gluing surfaces. 

There is an isomorphism $\chP_{g,n} \cong \chP'_{g,n}$, 
as we now explain.   
 For a marked surface $\Sigma$ in $\chP_{g,n}$, we obtain a surface $\Sigma'$ with $n$ boundary components $b_i$  by removing the discs $D_i$.  We define a coordinate $\theta_i$ on each $b_i$ by setting
\begin{equation} 
	\theta_i = \operatorname{Im} \log z_i. 
\end{equation}
Since $z_i$ is defined up to a phase, $\theta_i$ is defined up to a shift.  Further, the functions $\theta_i$ are all analytic and of range $2 \pi$ as desired.

Conversely, given a surface $\Sigma' \in \chP'_{g,n}$, we obtain a surface in $\chP_{g,n}$ by gluing a disc $\overline{D}_i = \{\abs{z_i} \le 1\}$ to the boundary component $b_i$ of $\Sigma$. This gluing identifies 
$\operatorname{Im} \log z_i$ with $\theta_i$.   
Because $\operatorname{Im} \log z_i$ and $\theta_i$ are, locally, the boundary values of holomorphic coordinates in a region of $\overline{D}_i$ 
and of $\Sigma'$, respectively, the glued surface is naturally holomorphic.

There is a map from $\chP'_{g,n}$ to the moduli space $\cM'_{g,n}$ of Riemann surfaces with boundary, given by forgetting the coordinates $\theta_i$. We will show that this map is homotopy equivalence. To show this, it suffices to show that the fibres are contractible (because $\chP'_{g,n}$ is a fibration over $\cM'_{g,n}$).
This means that we need to show that the space of possible choices of coordinate $\theta_i$ on each boundary component $b_i$ of $\cM'_{g,n}$ is contractible.  

Let us fix reference coordinates $\theta_i$ on each boundary component $b_i$, and write any other coordinate system as $\widetilde{\theta}_i = F_i(\theta_i)$.   Because each coordinate is defined up to a shift, we can assume that $F_i(0) = 0$.  For this new coordinate to have range $2\pi$ 
we must have $F_i(\theta_i + 2 \pi ) = F_i(\theta_i)+ 2\pi$.
 The compatibility with the orientation, and the fact that $F_i (\theta_i)$ must define a new coordinate system, tells us that $\partial_{\theta_i} F_i(\theta_i) > 0$. 

We need to show that the space of possible choices of functions $F_i(\theta_i)$ satisfying these constraints is contractible. The contracting homotopy is simply given by the one-parameter family 
\begin{equation} 
	F_i^\lambda(\theta_i) = \lambda \theta_i + (1-\lambda) F_i(\theta_i) \,,
\end{equation}
for $0 \le \lambda \le 1$.  This satisfies the same properties as $F_i(\theta_i)$, that is $\partial_{\theta_i} F_i^\lambda(\theta_i) > 0$, $F_i^\lambda(0) = 0$, and 
$F_i^\lambda(\theta_i + 2 \pi ) = F_i^\lambda(\theta_i) + 2\pi$.  
 Therefore $F_i^\lambda(\theta_i)$ defines a one-parameter family of real analytic coordinates connecting that given by $\theta_i$ to that given by $F_i(\theta_i)$.  

We have shown that $\chP_{g,n}$ is isomorphic to $\chP'_{g,n}$,  which is homotopy equivalent to the moduli space of Riemann surfaces with boundary. The latter is known to be homeomorphic to $\cM_{g,n} \times (\R_{> 0})^n$,
and so is homotopy equivalent to $\cM_{g,n}$. 

\subsubsection{Homology vanishing for $\cM_{g,n}$}

The homotopy equivalence between $\chP_{g,n}$ and $\cM_{g,n}$ shows that they have isomorphic homology groups. Next, we will show that  
$H_{6g-6+2n-1}(\cM_{g,n})$ 
vanishes except when $(g,n) = (0,4)$. 

By Poincar\'e duality, we can identify $H_{6g-6+2n-1}(\cM_{g,n})$  
 with $H^1(\overline{\cM}_{g,n}, \mbar_{g,n} \setminus \cM_{g,n} )$ (here we consider relative cohomology of the Deligne-Mumford space, relative to the complement of the locus of smooth Riemann surfaces). So it suffices to show that this $H^1$ cohomology group vanishes.   
 To do this, we first consider the exact sequence of relative cohomology groups
  \begin{equation} 
	  0 \to  H^0(\overline{\cM}_{g,n}) \to H^0(\partial \overline{\cM}_{g,n}) \to H^1(\overline{\cM}_{g,n}, \partial \overline{\cM}_{g,n}) \to   H^1(\overline{\cM}_{g,n}) \to \dots 
  \end{equation}
 It is known that $ H^1(\overline{\cM}_{g,n}) = 0$, and that $\partial \overline{\cM}_{g,n}$ is connected except for $(g,n) = (0,4)$.  This implies that, as desired,   
\begin{equation} 
	  H^1(\overline{\cM}_{g,n}, \partial \overline{\cM}_{g,n}) = 0 \,, 
	  \quad \hbox{for} \ \ (g,n) \neq (0,4)\,.  
\end{equation}
For $(g,n)=(0,4)$, the group $H_{1}(\cM_{0,4})$ does not vanish but  the 
$S_4$ invariant group $H_{1}(\cM_{0,4})^{S_4}$ does.  The moduli space $\cM_{0,4}$ is a $\CP^1$ with three points removed, as we can set the first three marked points to $0,1,\infty$ and the last marked point $z_4$ can be anywhere else on $\CP^1$.  

The first homology group of $\cM_{0,4}$ is two-dimensional, parametrized by the cycles where the last marked point $z_4$ moves around $0$ or moves around $1$. This is some two-dimensional representation of $S_4$, and we need to show it has no trivial subrepresentations.  To do this, it suffices to show that it has no trivial subrepresentations as a representation of the copy of $S_3$ in $S_4$ which permutes the first three marked point. As a representation of $S_3$, it is clear that this is the standard irreducible two-dimensional representation, which is the complement of the trivial representation in the three dimensional permutation representation. Therefore, there are no trivial subrepresentations. 
This completes the inductive proof of existence of the string vertices.

\subsection{Uniqueness of string vertices}
Uniqueness, up to BV canonical transformations, is proved by a similar inductive argument.   To understand this, we need to discuss the exponentiated form of the canonical transformation.

Let $\cV$ be a set of string vertices, and $\cW$ a collection $\cW_{g,n} \in \C_{6g-6+2n+1}(\chP_{g,n})$ of $S_n$-invariant   
singular chains which define an infinitesimal canonical transformation. 
We define  a family  $\cV(t)$ 
of string vertices such that $\cV(0)= \cV$ and 
 asking that they satisfy the differential equation
\be
\label{decV}
\frac{\mathrm{d}}{\mathrm{d} t}  \cV(t) =   \delta_\cW \cV(t)  \,.
\ee
where
\be
\delta_\cW \cV  =  \{ \cV\,, \cW \}  
	+ {\cal B}_\cW \, , \qquad  {\cal B}_\cW  \equiv \partial \cW
	+\hbar \Delta \cW \,, 
\ee
 is the infinitesimal canonical 
	transformation defined before, and we have introduced the symbol
	${\cal B}_\cW$ to denote the part of the transformation that is not linear
	on $\cV$.  

	If $\cV$ satisfies the master equation, then so does 
	$\cV(t)$ 
	for all $t$.  To see this, let us write the master equation 
for $\cV(t)$   
in the form
the form $M_\cV (t)= 0$ by defining
\be
M_\cV (t)  \equiv \partial \cV(t)  + \hbar \Delta \cV(t)   + \tfrac{1}{2} \left\{ 
 \cV(t), \cV(t) \right\}\,. 
\ee
Clearly $M_\cV (0)=0$ since the string vertices $\cV$ satisfy the master equation.
A short calculation using the differential equation~(\ref{decV}) shows that $M_\cV$
satisfies the equation
\be
\frac{\mathrm{d}M_{\cV}}{ \mathrm{d}t } =  \{  M_{\cV}(t) \,, \cW \} \,. 
\ee
This equation implies that if $M_\cV (0)=0$ all derivatives of $M_\cV(t)$ will
vanish at $t=0$.  This shows $M_\cV (t)$ vanishes at all times. 

We can write the solution of~(\ref{decV}) for the instantaneous vertices
by taking multiple derivatives, evaluating
at $t=0$ and writing the Taylor series.  This gives
\be
\cV(t) =     \cV + t \delta_\cW \cV  + \tfrac{1}{2} t^2 \{\delta_\cW \cV, \cW\} + \tfrac{1}{3!} t^3\bigl\{ \{\delta_\cW \cV, \cW\} , \cW\bigr\} + 
\cdots\ . 
\ee
We define the exponential of the canonical transformation
via  $ \Exp{ t\delta_{\cW} }\cV\equiv  \cV(t) $.  Note that the series solution does
not fit the naive expansion of the exponential in that we do not encounter nor
define iterated variations $\delta_{\cW} \delta_{\cW} \cV$.   The definition implies that
\be
 \Exp{ \delta_{\cW} }\cV \equiv   \cV +  \delta_\cW \cV  + \tfrac{1}{2}  \{\delta_\cW \cV, \cW\} + \tfrac{1}{3!}\bigl\{ \{\delta_\cW \cV, \cW\} , \cW\bigr\} + 
\cdots\ . 
\ee
With $\delta_\cW \cV = \{ \cV, \cW\}  + {\cal B}_\cW$ we have
\be
\label{exponential-in-detail}
\begin{split} 
 \Exp{ \delta_{\cW} }\cV = & \  \  \cV +  \{ \cV, \cW\}  + \tfrac{1}{2}  \{\{ \cV, \cW\}, \cW\} + \tfrac{1}{3!}\bigl\{ \{\{ \cV, \cW\}, \cW\} , \cW\bigr\} + 
\cdots\\
&   +  {\cal B}_\cW + \tfrac{1}{2}  \{ {\cal B}_\cW, \cW\} + \tfrac{1}{3!}
\bigl\{ \{ {\cal B}_\cW, \cW\} , \cW\bigr\} + \cdots \ . 
\end{split}
\ee
This equation will be useful below.

Now let us turn to the uniqueness of the string vertices.
Suppose that $\cV_{g,n}$, $\cV'_{g,n}$ are two sets of string vertices, both satisfying the master equation.  Our goal is to show inductively that there exists a sequence of $S_n$-invariant   
singular chains $\cW_{g,n} \in C_{6g-6+2n+1}(\chP_{g,n})$ such that
\begin{equation} 
	\Exp{ \delta_{\cW} } \cV = \cV'. \label{eqn:uniqueness} 
\end{equation}

To perform the induction it is useful to introduce a partial ordering on the collection of pairs $(g,n)$ of non-negative integers with $2g-2+n > 0$.  We say $(g',n') < (g,n)$ if $g'<g$, or if $g'=g$ and $n'<n$. 

The initial step of the induction is $(g,n) = (0,3)$, in which case $\cV_{0,3}$ and $\cV'_{0,3}$ are both $S_3$ invariant points in $\chP_{0,3}$.  Since this space is connected, we let $\cW_{0,3}$ be a $S_3$-invariant 
path connecting $\cV_{0,3}$ to $\cV'_{0,3}$. Viewing $\cW_{0,3}$ as a one-chain, we have
\begin{equation} 
	\partial \cW_{0,3} 
	= \cV'_{0,3} - \cV_{0,3}\,.  
\end{equation}
This implies that we have satisfied~(\ref{eqn:uniqueness}) to leading order
\begin{equation} 
\label{f-step}
	\left( \Exp{\delta_{\cW_{0,3}}} \cV \right)_{0,3} = \cV'_{0,3} \,,
\end{equation}
where for any chain $X$ we let $( X)_{g,n}$ 
denote the piece that belongs to $\chP_{g,n}$.  Equation~(\ref{f-step}) can be checked
using~(\ref{exponential-in-detail}) to evaluate the left-hand side.  
This means that the string vertices on the left and right 
of~(\ref{eqn:uniqueness}) agree for $(g,n) = (0,3)$.

Next, we assume that we have constructed $\cW_{g',n'}$ by induction 
for all $(g',n') < (g,n)$.   
We let $\cW_{<(g,n)}$ be the totality of the $\cW_{g',n'}$ that we have already constructed. 
By induction, we will assume that
\begin{equation} 
	\left( \Exp{\delta_{\cW_{<(g,n)}}} \cV \right)_{g',n'} =   \, \hbar^{g'}
	\cV'_{g',n'} \,,   
\end{equation}
for all $(g',n') < (g,n)$.  

To continue the induction, we need to find some 
$\cW_{g,n}$ such that   
\begin{equation} 
	\left( \Exp{\delta_{\cW_{<(g,n)}} + \hbar^g \delta_{\cW_{g,n}}  }
	 \cV \right)_{g',n'} = \,  \hbar^{g'} \, \cV'_{g',n'} \,,    
\end{equation}
for all $(g',n') \le (g,n)$.   
Changing the canonical transformation by adding $\hbar^g \cW_{g,n}$ does not affect the left hand side of this equation for $(g',n') < (g,n)$.  It only changes it for $(g',n') = (g,n)$. A bit of analysis   shows that
\begin{equation} 
\label{identity-for-Ws}
	\left( \Exp{\delta_{\cW_{<(g,n)}} + \hbar^g \delta_{\cW_{g,n}}  } \cV \right)_{g,n} =  \, \hbar^g\,  \partial \cW_{g,n}   
	+  \left( \Exp{\delta_{\cW_{<(g,n)}}} \cV \right)_{g,n} \,. 
\end{equation}
Indeed,  with $\cW = \cW_{< (g,n)} + \hbar^g \cW_{g,n}$, one can check using~(\ref{exponential-in-detail}) that the
contributions to a $(g,n)$ chain from $\cW_{g,n}$ arise only from ${\cal B}_\cW$,
and then only from the term with $\partial$.  In checking this it helps to note that for any $X_{g,n}$ and any $Y$, we have
$(\{ X_{g,n} , Y\} )_{\leq (g,n)} = 0$, namely, the left hand side only contains 
chains higher up in the ordering.   Similarly, for higher nested objects
$(\bigl\{ \{ X_{g,n} , Y\}, Z \bigr\}  )_{\leq (g,n)} = 0$.

It is useful to define   
\begin{equation} 
	\cV'' := \Exp{\delta_{\cW_{<(g,n)} }} \cV \,.
\end{equation}
With this, the induction assumption above states
that $\cV''_{g',n'} =  \cV'_{g',n'}$ for all $(g',n') < (g,n)$.
 To continue the induction and on account of~(\ref{identity-for-Ws}),  
the chain $\cW_{g,n}$ is required to satisfy  
\begin{equation} 
	\partial \cW_{g,n} = 
	\cV'_{g,n} - \cV''_{g,n} \,.  
	\label{eqn:inductive_step_w} 
\end{equation}
The right hand side is a singular chain in $C_{6g-6+2n}(\chP_{g,n})$.  This equation can only have a solution if the right hand side is in the kernel of the boundary operator $\partial$.
To see this, we note that the master equation expresses $\partial \cV'_{g,n}$ and $\partial \cV''_{g,n}$ in terms of $\cV'_{g',n'}$ and $\cV''_{g',n'}$,
respectively,  
 for $(g',n') < (g,n)$. These spaces are the same by the induction assumption, so that, as expected,  
\begin{equation} 
\label{s-eqndpp}
	\partial \, (\cV''_{g,n} - \cV'_{g,n} )= 0. 
\end{equation}
To show that there exists a solution $\cW_{g,n}$ to  \eqref{eqn:inductive_step_w}, it suffices to show that the homology group $H_{6g-6+2n}(\chP_{g,n})$ is zero (this, of course implies it also vanishes for $S_n$-invariant chains).  This is the case as long as $(g,n) \neq (0,3)$. The point is that $\chP_{g,n}$ is homotopy equivalent to $\cM_{g,n}$, and $\cM_{g,n}$ is a non-compact orbifold of dimension $6g-6+2n$.  By Poincar\'e duality, $H_{6g-6+2n}(\cM_{g,n})$ is isomorphic to $H^0(\mbar_{g,n}, \mbar_{g,n} \setminus \cM_{g,n} )$, the cohomology of the Deligne-Mumford compactification relative to the complement of the open subset of smooth surfaces. This cohomology group vanishes, except in the case $(g,n) = (0,3)$.  
This completes the proof of uniqueness (up to canonical transformations) of the string vertices.

\subsection{Representing the fundamental homology class of moduli space} 
We have not, however, addressed an important part of the story.  Given a collection $\cV_{g,n}$ of string vertices, we let $\cF_{g,n}$ be the corresponding chains built using Feynman diagrams with $\cV$ as the vertices.  We need to show that when we push forward the chain $\cF_{g,n}$ to $\cM_{g,n}$, we find a representative of the fundamental homology class\footnote{Although $\cM_{g,n}$ is strictly speaking an orbifold and not a manifold (because the mapping class group action on Teichm\"uller space is not free)  
 there is no difficulty in defining the fundamental class. } of  $\cM_{g,n}$. 

To understand this, we need a few more details on the construction of $\cF_{g,n}$. We take as our propagator the 
flat cylinder of fixed radius  and arbitrary length $\tau$. According to the usual string field theory Feynman rules, when we glue in the cylinder, we allow a ``twist''. This means that adding a propagator increases the dimension of a chain by two: one for the length $\tau$  and one for the twist parameter.

Let us include into our propagator the infinite cylinder $\tau = \infty$.  Conformally, having a cylinder of infinite length is equivalent to having a cylinder of finite length but with a circle of radius zero in the middle. We will interpret the infinite length cylinder in this way, as two very long ``cigars'' meeting at their tips.  In terms of algebraic geometry, we can view it as the nodal curve $zw = 0$.   At infinite length, the twist parameter in the Feynman diagrams is irrelevant, as we can rotate each cigar independently.  This should be familiar: when we glue two Riemann surfaces along a common boundary to produce a smooth surface, we need to specify the twist to perform the gluing. But when we glue two surfaces with marked points to produce a surface with a nodal singularity, no such choice is necessary.

If we include this infinite cylinder in our propagator, then we see that $\cF_{g,n}$ is a chain in the space $\widehat{\overline{\cP}}_{g,n}$: this is 
the space consisting of possibly nodal surfaces  in the Deligne-Mumford compactification $\overline{\cM}_{g,n}$ equipped with coordinates around the punctures, defined up to rotation.  

By forgetting the coordinates around the punctures, there is a map 
\begin{equation} 
	\pi :  \widehat{\overline{\cP}}_{g,n} \to \overline{\cM}_{g,n}.
\end{equation}
There is a corresponding map $\pi_\ast$ on singular chains, and our goal is to show that $\pi_\ast \cF_{g,n}$ is a representative for the fundamental homology class of $\overline{\cM}_{g,n}$.

To show this, we will first show that the chain $\cF_{g,n}$ is closed. The boundary $\partial \cF_{g,n} = 0$ has three contributions:
\begin{enumerate} 
	
\item Contributions from the boundary of the vertices $\cV_{g',n'}$ making up $\cF_{g,n}$.
	
\item Contributions from 
surfaces where the length $\tau$ of a propagator is zero.

\item Contributions from surfaces where the length $\tau$ of the propagator
is $\infty$.

\end{enumerate}
The first two types of contributions cancel each other exactly, because of the master equation satisfied by the vertices $\cV$.  The key point is that the propagator at $\tau = 0$ performs precisely the ``twist-gluing'' defining the operators $\Delta$ 
and $\{ \cdot \, , \, \cdot \}$ 
in the master equation. The third contribution, that from $\tau = \infty$, vanishes as well. This is because the twist parameter in the gluing is not present at $\tau = \infty$, so that this locus is not a boundary, but rather of codimension $2$.    
This completes the proof that $\cF_{g,n}$ is closed. 

Since 
$\cF_{g,n}$ is a cycle (i.e.\ closed chain) in $C_{6g-6+2n}(\widehat{\overline{\cP}}_{g,n})$,  its pushforward $\pi_\ast \cF_{g,n}$ to $\overline{\cM}_{g,n}$ is 
a cycle in $C_{6g-6+2n}({\overline{\cM}}_{g,n})$.  
Because $\overline{\cM}_{g,n}$ is a 
compact orbifold,\footnote{The fact that $\cM_{g,n}$ is an orbifold is important here because Poincar\'e duality, with rational or real coefficients, holds for orbifolds with finite stabilizer groups.}  
Poincar\'e duality tells us that $H_{6g-6+2n}(\overline{\cM}_{g,n}) = \R$.  It follows that there is some constant $c_{g,n} \in \R $ such that
\begin{equation} 
	[\pi_\ast \cF_{g,n}] = c_{g,n} [\mbar_{g,n}]. 
\end{equation}
We need to show that $c_{g,n} = 1$.

If the cycle $\cF_{g,n}$ was itself a manifold or an orbifold, then the statement that $[\pi_\ast \cF_{g,n}] = [\mbar_{g,n}]$ would be equivalent to saying that the map $\cF_{g,n} \to \mbar_{g,n}$ is of degree one.  
Being of degree one is a condition that is local on $\mbar_{g,n}$: to check it, it suffices to find a point in $\mbar_{g,n}$ such that the map $\cF_{g,n} \to \mbar_{g,n}$ is an isomorphism near that point.  

Since $\cF_{g,n}$ is not necessarily represented by an orbifold, we can not apply this argument directly. The constraint that $\cF_{g,n}$ is represented globally by an orbifold is not strictly necessary, however. The argument continues to apply if we can find a small region in $\mbar_{g,n}$ such that over this region, $\cF_{g,n}$ is represented by an orbifold which projects isomorphically onto $\mbar_{g,n}$.  We will see that this is the case. 

Fix a trivalent graph $\gamma$ with $g$ loops and $n$ external lines, and use it to build a totally degenerate nodal surface $\Sigma_\gamma$ 
in $\mbar_{g,n}$,   by gluing together spheres with three points according to $\gamma$ 
and having all propagators with $\tau = \infty$.  
We can find a small neighbourhood $U \subset \mbar_{g,n}$ of $\Sigma_\gamma$ such that, on the subset $\pi^{-1} (U) \subset \widehat{\overline{\cP}}_{g,n}$, the chain $\cF_{g,n}$ is represented entirely by the trivalent Feynman graph $\gamma$ with the vertex $\cV_{0,3}$. 

We see that in this neighbourhood of $\Sigma_\gamma$, $\cF_{g,n}$ is represented  by an orbifold, whose coordinates are the length and twist parameters of the propagators, where all the length parameters are very large.  The orbifold group is the automorphism group of the graph $\gamma$.  Evidently, this orbifold 
projects isomorphically onto a neighbourhood of $\Sigma_\gamma$: every surface obtained from smoothing $\Sigma_\gamma$ is specified uniquely by its length and twist parameters, up to the action of the automorphism group of $\Sigma_\gamma$.   Because, in a neighbourhood of $\Sigma_\gamma$, $\cF_{g,n}$ is represented by a manifold which projects isomorphically onto $\mbar_{g,n}$, 
we conclude that, as desired,  
\begin{equation} 
	[\pi_\ast \cF_{g,n}] = [\mbar_{g,n}]  \,.
\end{equation}
This concludes the proof that the string vertices $\cV$ constructed to satisfy
the master equation will built through Feynman graphs a chain $\cF_{g,n}$
that pushed to $\mbar_{g,n}$ represents its fundamental homology class.

\section{Preliminaries for the hyperbolic construction}

In this section we prepare the ground for the definition of string vertices
and the proof that they satisfy the geometric master equation.  We begin
by discussing the relevant moduli space of hyperbolic metrics with
geodesic boundaries and explain how to pass to the moduli space of surfaces
with punctures.  We then consider the collar theorems of hyperbolic
geometry that will ensure the consistency of the vertices upon gluing
across boundaries. 

\subsection{Moduli spaces of bordered and punctured surfaces}

Consider an orientable surface $S$ of genus $g$ with 
$n$ boundaries.  We let ${\cal T}_{g,n} (S)$ denote the Teichmuller
space of (marked) hyperbolic metrics on the surface $S$ where the
boundaries are geodesics.  By the uniformization
theorem this is also the Teichm\"uller space of (marked) 
complex structures on the
surface $S$, namely the set of all (marked) Riemann surfaces of genus
$g$ with $n$ boundaries.  

The space ${\cal T}_{g,n} (S)$ is very large as it contains 
metrics with all values of the lengths of the boundary components.  
Let us restrict ourselves to ${\cal T}_{g,n,L} (S)$ defined as the subspace of 
${\cal T}_{g,n}(S)$
where all boundaries have length $L$.  Let $\Gamma(S)$ denote the mapping 
class group of $S$: the quotient of the set of all orientation
preserving diffeomorphisms by the set of all diffeomorphisms connected
to the identity (all diffeomorphisms must preserve the boundaries).  We then have the moduli space ${\cal M}_{g,n, L}$
\be
{\cal M}_{g,n,L} \equiv  {\cal T}_{g,n,L} (S)/ \Gamma(S) \,. 
\ee
This is a moduli space of Riemann surfaces of genus $g$ with $n$ boundaries. 
The surfaces here come equipped with hyperbolic metrics (defining the
complex structure) and have geodesic
boundaries of length $L$. 

\begin{figure}[!ht]
	\centering
	\fd{9.0cm}{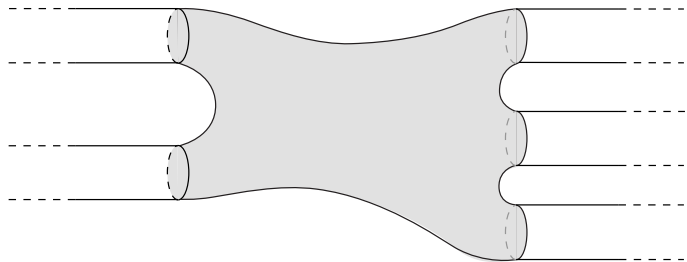}  
	\caption{A hyperbolic surface $\tilde\Sigma_{g,n}$ with $n$ boundary
	geodesics of length $L$ becomes a surface $\Sigma_{g,n}\in \chP_{g,n}$
	by attaching flat semi-infinite cylinders of circumference $L$
	at each boundary.}
	\label{fig:nhole}
\end{figure}

The surfaces in ${\cal M}_{g,n,L}$ have boundaries, but for string 
vertices we need surfaces with marked points and local coordinates
about them.\footnote{An alternative description of vertices 
uses surfaces with
{\em parameterized} boundaries.  Here the parameterization would
be by the length function provided by the metric.  Both descriptions
are equivalent (see section~\ref{pro-the-hom-equ}). }  There is a canonical way to obtain such surfaces: we
attach flat semi-infinite cylinders of circumference $L$ at each boundary.
The gluing is done isometrically, and the metric is continuous at the seam. 
Each semi-infinite cylinder is conformal to a punctured disk and thus
introduces automatically the puncture and the local coordinate:
the coordinate $z$ in which the disk is $|z| \leq 1$ and the puncture
is at $z=0$.   
We think of this as the operation `$\hbox{gr}'_\infty$' of  {\em grafting} 
those cylinders on surfaces:  
\be
\hbox{gr}'_\infty :  \ {\cal M}_{g,n,L} \to \chP_{g,n}\,. 
\ee
Thus for $\tilde \Sigma_{g,n} \in {\cal M}_{g,n,L}$ we obtain
a surface $\Sigma_{g,n} = \hbox{gr}'_\infty \tilde \Sigma_{g,n} \in \chP_{g,n}$:
The grafting map is illustrated in Figure~\ref{fig:nhole}.

Using the projection map $\pi: \chP_{g,n} \to \cM_{g,n}$ 
we now consider the composition $\pi \circ \hbox{gr}'_\infty$
\be
\hbox{gr}_\infty  \equiv \pi \circ \hbox{gr}'_\infty :  \ {\cal M}_{g,n,L} \to \cM_{g,n}\,. 
\ee
The map gr${}_\infty$ can be shown to be a homeomorphism thus, in particular,
a one to one, onto map.  This claim follows from
 a result by Mondello~\cite{mondello}.  Theorem 5.4 ~in\cite{mondello}
 establishes that
 with ${\cal T}_{g,n}$ the Teichm\"uller
 space of $n$-punctured genus $g$ surfaces,  
 the grafting map $\hbox{gr}_\infty :   {\cal T}_{g,n,L} \to  {\cal T}_{g,n}$
 is a mapping-class group equivariant homeomorphism.   A key tool in
 the analysis of~\cite{mondello} was developed by Scannell and Wolf~\cite{scannell},
 who showed that the effect of grafting a finite cylinder onto a geodesic induces
 a homeomorphism of Teichm\"uller space.

\subsection{Collar theorems for hyperbolic metrics}

Given a surface $S$ with a hyperbolic metric and a simple closed
geodesic $\gamma$ the {\em collar}  ${\cal C}(\gamma)$ of 
width $w$ about $\gamma$
is the set of all points whose distance to $\gamma$
does not exceed $w/2$:
\be
{\cal C}(\gamma) = \bigl\{ p \in S \bigl| \,  \hbox{dist} (p, \gamma) \leq 
\tfrac{w}{2} \bigr\} \,.
\ee
If the geodesic $\gamma$ is a boundary geodesic, ${\cal C}(\gamma)$ 
is a half-collar of width
$w/2$.
The relevance of properly chosen 
collars is clear when we have a collection of 
simple closed geodesics $\gamma_i$ of length $\ell(\gamma_i)$ 
 that do not intersect.  Consider collars of width $w(\gamma_i)$
for interior geodesics and half-collars of width $\tfrac{1}{2}w(\gamma_i)$
for boundary geodesics.  The widths of the collars are chosen so that 
\be
\sinh \bigl(\tfrac{1}{2} w(\gamma_i)\bigr)\, 
\sinh \bigl(\tfrac{1}{2} \ell(\gamma_i)\bigr)\,  = \, 1\,.
\ee
It is a well know result of hyperbolic
geometry that such collars are disjoint on the surface~\cite{buser}. 

Define now the length $L_*$:
\be
L_* = 2\, \hbox{arcsinh} \, 1 = 2 \log (1+ \sqrt{2}) = \log ( 3+ 2\sqrt{2}) \simeq 1.76275\,. 
\ee
For geodesics of length $L_*$ the associated collar width $w_*$ is
in fact equal to $L_*$:
\be
w_* = L_*\,.
\ee
Consider now half-collars associated with boundary geodesics
of length $L \leq  L_*$   
and let  $w_L/2$ denote the width of the half-collars.
We then have: 
 \be
\sinh \bigl(\tfrac{1}{2} w_L\bigr)\, 
\,  = \, {1\over \sinh \bigl(\tfrac{1}{2} L\bigr)} \geq   {1\over \sinh \bigl(\tfrac{1}{2} L_*\bigr)}  =  1\,,
\ee
so that $\sinh \bigl(\tfrac{1}{2} w_L\bigr) \geq 1$.  Since 
$\sinh \bigl(\tfrac{1}{2} L_*\bigr) = 1$ and sinh grows monotonically 
for positive arguments,  we conclude that  
$\tfrac{1}{2} w_L \geq  \tfrac{1}{2} L_*$. Finally,  since $L_* > L$, 
\be
\label{width-length}
    \tfrac{1}{2} w_L \geq  \tfrac{1}{2} L \,, \quad \hbox{for} \ \ L< L_* \,.   
\ee
In Figure~\ref{fig:3hole} we show the half-collars associated with a genus
zero surface with three boundaries.

Another useful property follows from the collar theorem.
Given a simple closed geodesic~$\gamma$ and another geodesic $\delta$
(not necessarily simple) that intersects $\gamma$ transversely, 
their lengths satisfy~\cite{buser} (Corollary 4.1.2)  
\be
\label{xcrossing1}
\sinh \bigl(\tfrac{1}{2} \ell(\gamma)\bigr)\, 
\sinh \bigl(\tfrac{1}{2} \ell(\delta)\bigr)\,  > \, 1\,.
\ee
We can now immediately conclude that:

\noindent
{\bf Claim 1.} A simple geodesic of length $L \leq  L_*$ 
cannot intersect another geodesic (not necessarily simple) 
of length $L' \leq L_*$. 
These must be disjoint.
\begin{figure}[!ht]
	\centering
	\fd{5cm}{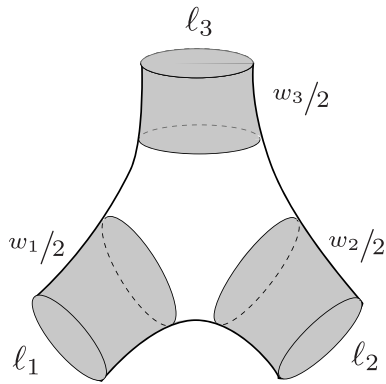}
	\caption{Hyperbolic pants with the half-collars associated
	with the boundary geodesics. When $\ell_1= \ell_2 = \ell_3=L < L_*$
	the half collars shown are all of length greater than $L/2$.}
	\label{fig:3hole}
\end{figure}

\section{Hyperbolic string vertices}
\label{hyp-str-ver}

In this section we introduce  
a set of hyperbolic string vertices and demonstrate that they provide
 an exact solution of the master equation~(\ref{m-eqn}).  
 We also describe in some detail the simplest vertex $\cV_{0,3}$,
 corresponding to a sphere with three punctures.   We discuss what
 is entailed in finding a concrete description of the genus one, once-punctured 
 vertex $\cV_{1,1}$.
 
 \subsection{Hyperbolic vertices as systolic subsets} 
 
 We first define
 a subset $\tilde \cV_{g,n}(L)$ of the moduli space $\cM_{g,n,L}$
 of genus $g$ surfaces with $n$ boundaries of length $L$:  
\be
\widetilde\cV_{g,n}(L) \equiv  \ 
\bigl\{\  \tilde \Sigma \in \cM_{g,n,L} \bigl| \   
 \hbox{sys} (\tilde \Sigma) \geq L \bigr\} \,.  
 \ee
Here  $\hbox{sys} (\tilde\Sigma)$ is the systole of $\tilde\Sigma$: the length of the shortest non-contractible
 closed geodesic in $\tilde \Sigma$ which is not a boundary component.   
 Since the boundary geodesics have length $L$,  
  the above sets include all surfaces in $\cM_{g,n,L}$
 that have no geodesic of length less than $L$.  
 This definition applies for all vertices listed in~(\ref{formal-sum}),  namely
 $n\geq 3$ for $g=0$,  $n\geq 1$ for $g=1$, and $n\geq 0$ for $g\geq 2$.
For $n=0$, the space ${\cal M}_{g,0,L}$ is just ${\cal M}_{g,0}$.   
 
The string vertices $\cV_{g,n}(L)$ are now obtained by applying $\hbox{gr}'_\infty$
to the above sets, namely, by grafting the semi-infinite cylinders to turn 
the boundaries into punctures with local coordinates: 
\be
\cV_{g,n}(L) \equiv  \hbox{gr}'_\infty \, \bigl(  \widetilde\cV_{g,n}(L) \bigr) \,. 
\ee

\noindent
{\bf Theorem 1:} {\em The sets $\cV_{g,n}(L)$ with $L \leq L_*$ solve the master equation~(\ref{m-eqn}).}

\noindent
{\bf Proof.}
We must now show that the above string vertices, defining $\cV$ 
as the formal sum~(\ref{formal-sum}), solve the equation:
\be
\label{m-eqn-again}
\partial \cV  =  - \hbar \Delta \cV  - \tfrac{1}{2} \{ \cV\,, \cV \} \,.
\ee
We first show that the surfaces appearing in the 
boundary of $\cV$ are contained on the
right-hand side.  The boundary of $\cV$ arises when a simple closed 
interior geodesic becomes of length~$L$. 
The geodesic cannot be non-simple since all such geodesics
are at least of length $2L_*$ (see~\cite{buser}, Theorem 4.2.2).  
Note that two  intersecting closed geodesics cannot reach length 
$L$ simultaneously either
because of Claim 1.   
When more than one simple
closed geodesic becomes of length $L$ we are simply at a lower codimension 
set on the boundary.

Suppose we cut the surface $S$ at the  length $L$ geodesic, 
obtaining the surface $S_c$.  Then
either $S_c$ is two disconnected surfaces, or $S_c$ remains 
connected.  If $S_c$ is disconnected each surface is in a string
vertex, for none has a geodesic shorter than $L$, 
and all boundaries are geodesics of length $L$.  
Then $S$ arises from $\{ \cV\,, \cV \}$. 
If $S_c$ remains 
connected, $S_c$ has two more boundaries
than $S$, and is contained in a string vertex, having no 
geodesic shorter than $L$ and boundaries of length $L$. Then $S$ arises from 
$\Delta$ acting on $S_c$.  This proves that the boundary of $\cV$ 
is indeed contained on the right-hand side.

Now we must show that the sewing operations on the right-hand side of
(\ref{m-eqn-again}) acting on surfaces belonging to string vertices
give us surfaces on the boundary of the string vertices. 
Consider first the case when we sew together boundary geodesics
on two separate surfaces $S_1$ and $S_2$, each from a string vertex.

At the seam the new surface has a geodesic of length exactly~$L$.
The new surface must be checked to have systole $L$.
No geodesic totally contained in $S_1$ or in $S_2$ is shorter than $L$, 
so the only concern
is that a geodesic $\gamma$ that runs on both surfaces is shorter than $L$.  
Such curve would have to intersect the simple closed curve at the seam
but this is impossible due to Claim~1. 

In fact, the condition $L \leq L_*$ used for Claim 1
 is not needed for these curves to work out. 
Any short geodesic $\gamma$ running on both 
surfaces has a piece of length $\ell_1$ in
$S_1$ and a piece of length $\ell_2$ in $S_2$, with $\ell_1 + \ell_2< L$.
The geodesic cuts the seam into two pieces of length $a_1$ and $a_2$ with
$a_1 + a_2 = L$.   At least one of $\ell_1, \ell_2$, call it $\ell_i$, 
 must be smaller than $L/2$.  And at least one of $a_1, a_2$, call it 
 $a_j$, satisfies $a_j \leq L/2$.  The closed curve $\tilde \gamma$ 
 formed by $\ell_i$ 
 followed by $a_j$ would be smaller than $L$ and would be contained
 in $S_i$.  Moreover $\tilde \gamma$ is
  a non-contractible curve, because otherwise,
 $\gamma$ itself would be homotopic to a curve in just one of the two
 surfaces.  In summary, $\tilde \gamma$ is a non-contractible closed
 curve on $S_i$ that is shorter than $L$. This is impossible, and proves the claim. 
 
If we sew two boundary geodesics on a single surface belonging to a string
vertex, we create a new handle with a geodesic
of length $L$ along the seam.  
Any new potentially short curve must traverse this collar and intersect
the seam, which is ruled out by Claim~1. 
Thus the surfaces on the right-hand side
of (\ref{m-eqn-again}) are in $\partial \cV$.   In this case the collar
was required.   This concludes the proof that the chosen
string vertices satisfy the geometric master equation.  \hfill $\square$
 
\noindent
{\em Comments:}    
\begin{enumerate}   
\item Since the map $\hbox{gr}_\infty :   {\cal M}_{g,n,L} \to \cM_{g,n}$ is a homeomorphism
it follows that the string vertices define a piece of a section over the moduli space $\cM_{g,n}$.

\item  The vertices cover what is usually called the `thick' part of the moduli
space, the part including all surfaces with systole greater than or equal to
some positive number $\epsilon$.  The vertices are a compact subset of
moduli space, by Mumford's compactness criterion~\cite{mumford}. 

\item All string vertices are non-empty sets.  This is 
nontrivial: for a choice of $L$ the systole is a function on
$\cM_{g,n,L}$ that is bounded above.  Over the moduli
space $\cM_{2,0}$, for example, the systole has a maximum at the Bolza surface 
with value $2\,  \hbox{arccosh} (1 + \sqrt{2}) \simeq 3.057$~\cite{jenni,parlier}.  
This surface would be included in 
any $\cV_{2,0}(L)$ since $L \leq L_*\simeq 1.7628$.

\end{enumerate} 

\subsection{The three-string vertex}

The lowest-dimensional subset in ${\cal V}$ is ${\cal V}_{0,3}(L)$.  This subset includes
one surface, a sphere with three punctures.  By definition,  $\tilde\cV_{0,3}(L)$ is the hyperbolic pants with 
three geodesic boundaries all of length $L$.  There is just one surface
in the corresponding moduli space.  To this surface we again
attach semi-infinite cylinders to obtain $\cV_{0,3}(L)$ (see Figure~\ref{fig:3hole}).
\begin{figure}[!ht]
	\centering
	\fd{5cm}{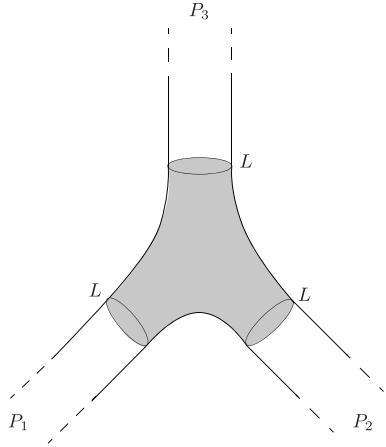} 
	\caption{The surface ${\cal V}_{0,3}(L)$ is the three-punctured
	sphere obtained by attaching three semi-infinite cylinders of circumference 
	$L$ to the
	geodesic boundaries of the symmetric hyperbolic pants with boundary lengths all set equal to $L$. }
	\label{fig:3hole}
\end{figure}

\begin{figure}[!ht]
	\centering
	\fd{12cm}{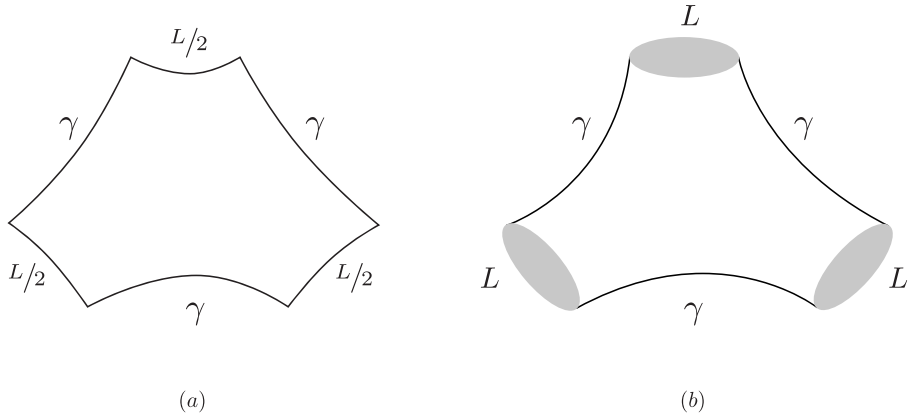}  
	\caption{(a) A hexagon with geodesic edges that meet orthogonally
	at each vertex.  This is a special vertex with three edges of length $L/2$ 
	and three edges of length $\gamma$.  (b) Gluing two such hexagons
	across the length $\gamma$ edges we obtain a pants diagram, the
	string vertex with three equal boundaries that are geodesics of length $L$. }
	\label{fig:2hex}
\end{figure}

The symmetric pants with boundaries of length $L$
 is constructed from gluing two copies
of a  geodesic hexagon with
sides lengths 
$\tfrac{L}{2}, \gamma , \tfrac{L}{2}, \gamma ,\tfrac{L}{2}, \gamma$, as shown in 
Figure~\ref{fig:2hex}.
The length $\gamma$ is the geodesic length between any pair of boundaries.
Hexagon identities (see~\cite{buser}, Theorem 2.4.1)  tell us that $\gamma$
is given by
\be
\cosh \gamma =  { \cosh \tfrac{L}{2}  \over \cosh \tfrac{L}{2}  -1} \,. 
\ee
This shows that as $L \to \infty$, the distance $\gamma$ between geodesic
boundaries 
goes to zero: $\gamma \simeq  2 e^{-L/4}$.  This is as expected because
the area $A_{0,3}$ of the surface is fixed.  In fact, from Gauss Bonnet we have
$A_{g,b} = 2\pi (2g-2 + b)$ with $g$ and $b$ the genus and the number of boundaries.
This gives $A_{0,3}= 2\pi$ for the hyperbolic $\cV_{0,3}(L)$, independent of the value of $L$.  With the area fixed and the boundaries becoming infinitely long, the
distance between the boundaries is going to zero. 
 
\begin{figure}[!ht]
	\centering
	\fd{11cm}{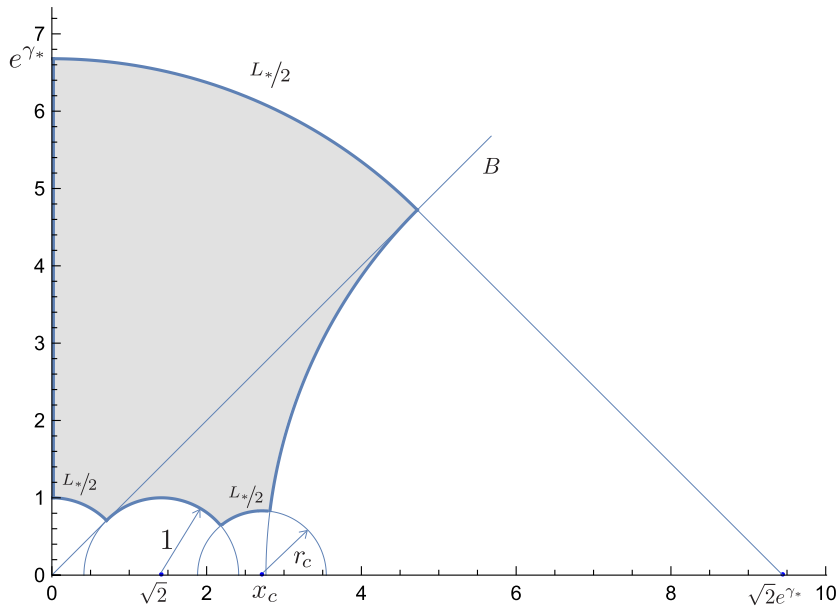}
	\caption{The hyperbolic vertex $\cV_{0,3}(L_*)$ is constructed
	by gluing together two hyperbolic hexagons with alternating edges
	$L_*$ and $\gamma_*$ across the $\gamma_*$ edges.  The hyperbolic
	hexagon is shown above, as the shaded area.  The line $B$ is at exactly
	at $\pi/4$ angle, and the circular arcs from the imaginary axis to $B$ are
	two sides of length $L_*/2$.  All other sides are of length $\gamma_*$.}
	\label{fig:HVert}
\end{figure}

To become more familiar with the hyperbolic vertex we consider $\cV_{0,3}(L_*)$,
the vertex for the largest possible value of $L$.  In this case
 $\cosh \tfrac{L*}{2}  = \sqrt{2}$ and $\gamma_*$ is readily determined:
\be
\cosh \gamma_* =  2 + \sqrt{2}\,,  \quad  \gamma_* = \log \Bigl( 2 + \sqrt{2} + \sqrt{5 + 4 \sqrt{2} }\, \Bigr) \simeq  1.89892\,. 
\ee
Since the collar width $w_*= L_* \simeq 1.76275< \gamma_*$, the half collars around the boundary geodesics, each of width $w_*/2$, do not touch.  This is
as it should.   The vertex $\cV_{0,3}(L_*)$ is built by gluing two hexagons, 
each of sides $\tfrac{L_*}{2}, \gamma_* , \tfrac{L_*}{2}, \gamma_* ,\tfrac{L_*}{2}, \gamma_*$.  Each hexagon can be displayed in the upper half-plane $\mathbb{H}$, with standard metric $ds = |dz|/y$.
We do this in Figure~\ref{fig:HVert}.  The construction follows the method used to
show that hyperbolic hexagons with prescribed lengths for three non-consecutive edges exist (see~\cite{buser}, Section 1.7).  Two of the sides of length $L_*/2$
appear as the arcs from the imaginary axis to the line $B$, that has slope one.
There is a circle centered at $\sqrt{2}$ with radius one, and a circle at $x_c$ with
radius $r_c$, given by
\be
x_c = {1\over 4} \Bigl( 2 + 3\sqrt{2} + \sqrt{2} \sqrt{5 + 4\sqrt{2}} \ \Bigr) \simeq 2.715\,, \quad r_c^2 =  
{1\over 2} \sqrt{  \sqrt{2}+  \sqrt{ 2\sqrt{2}-1} \ }\ \simeq 0.8316\,. 
\ee
The vertical edge runs from $i$ to $i e^{\gamma_*}\simeq i \, 6.679$. There 
is a circle centered at $\sqrt{2}e^{\gamma_*} \simeq  9.445$ with
radius~$e^{\gamma_*}$; its intersection with the real axis is
close but does not coincide with $x_c$.\footnote{It is possible to represent the 
full hyperbolic vertex, after two cuts, as a octagon in 
$\mathbb{H}$ (see~\cite{hubbard}, section 3.8).} 

\noindent
\underline{Comments on the vertex $\cV_{1,1}(L)$.} 

We wish to illustrate what is involved in giving an explicit construction
of the systolic set $\cV_{1,1}(L)$.  We begin with
the Teichm\"uller space ${\cal T}_{1,1,L}$ 
of genus one surfaces with one
geodesic boundary of length $L$.  This space can be described with Fenchel-Nielsen
coordinates.  We take a set of pants with boundary lengths $L,L', L'$ and glue 
the length-$L'$ boundaries 
together with twist parameter $\alpha$.  
As $L' \in [0,\infty]$ and $\alpha\in (-\infty, \infty)$, the full space ${\cal T}_{1,1,L}$ is generated.  

Two conditions select the subset $\cV_{1,1}(L)$: systole $L$ and restriction to
 inequivalent surfaces under the action of the mapping
class group.  Since surfaces with $\alpha$ and $\alpha+1$ are the same Riemann surface, we can restrict ourselves to $\alpha \in [0, 1]$.   Moreover, since $L'$ is the
length of a nontrivial closed geodesic, so we must take $ L' > L$.   This, however,
does not guarantee that the systole is $L$; other geodesics can become short.
In particular, when $\alpha = {1\over 2}$ (the case when gluing turns the geodesic between the two length-$L'$ boundaries into a closed geodesic) and $L'$ is  large,  there is a geodesic shorter than $L$.  The systolic
condition defines a nontrivial subregion
in the space $L' > L, \alpha \in [0,1]$.  Within this region, one must determine the set of inequivalent 
surfaces. This is challenging since the mapping class group does
not have a simple action on Fenchel-Nielsen coordinates.\footnote{Perhaps the results of L. Keen~\cite{keen} could help in this step.}

\section{The Feynman region} 

In this section we consider the Feynman diagrams formed with
the hyperbolic string vertices defined in section~\ref{hyp-str-ver}.  We will
explain how the obvious metric on the Riemann surfaces is
hyperbolic at the vertices and flat on the cylinders that represent the
propagators.  We discuss the mathematical framework relevant
to these {\em Thurston} metrics.  This is the theory
of complex projective structures and measured laminations, which
suggests these metrics are rather natural (for a review of these topics
see~\cite{dumas-review}).  
We are unable to determine if the Feynman graphs
provide a section of $\chP_{g,n}$ over $\cM_{g,n}$ but discuss
partial results in the literature that indicate the affirmative 
answer is quite plausible.

\medskip
When the vertices are glued directly across 
 length $L$ boundaries
the resulting surface remains hyperbolic; that was in fact
a  requirement for the vertices to satisfy the master equation.  
If a propagator is added, however,
an annulus is grafted: the boundaries of the annulus are
attached to boundaries on the surfaces (on two different
surfaces or on the same surface).  Once we insert the annulus
there is no direct way to get a hyperbolic metric on the whole surface.
If we put a hyperbolic metric on the annulus, the total metric is 
discontinuous at the seams, thus not hyperbolic. 
The string field theory indicates that the cylinder, of circumference $L$ 
must be added with all values $t > 0$ of the height, and with all values
$0 \leq \theta < 2\pi$ of the twist angle~$\theta$.  
This range of the twist takes into account that in moduli space we get the same surface  for $\theta=0$ and $\theta= 2\pi$.   In the Feynman region
we generally include several propagators, so we will have a 
collection $(t_i, \theta_i)$ of parameters, with $i= 1, \cdots, N_p$, with 
 $N_p$ the number of propagators.

\begin{figure}[!ht]
	\centering
	\fd{10cm}{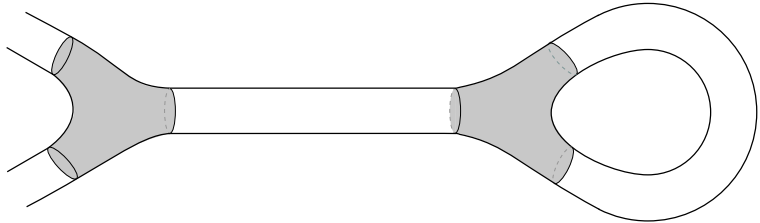}  
	\caption{A Feynman diagram gives Riemann surfaces with metrics that
	are hyperbolic on the string vertices (shown shaded)  and flat on the propagators.}
	\label{fig:fdiag}
\end{figure}

In string field theory it is natural 
to introduce such annuli
as finite-length flat cylinders of circumference $L$ that are grafted
into the surface.  In this way
the metric on the whole surface is partially hyperbolic on the vertices
and flat on the grafting cylinders (Figure~\ref{fig:fdiag}).  The metric is $C^1$ continuous but
not smooth: the curvature changes discontinuously across seams
from $K= -1$ on the hyperbolic part to
$K=0$ on the cylinder.  This metric is called a {\em Thurston metric 
on the surface}, and arises naturally through an intrinsic definition~\cite{tanigawa}
reviewed below.

To begin, consider a complex projective $\mathbb{CP}^1$ 
structure on a compact surface $S$:  an atlas of charts valued on
 $\mathbb{CP}^1$ with M\"obius transition functions. We call 
 ${\cal P}(S)$ the set of (isotopy classes of) marked complex projective
 structures.  Since M\"obius maps are holomorphic there is a projection
 map to Teichm\"uller space $\pi:  {\cal P}(S) \to {\cal T}(S)$.  It is known that one can identify
 ${\cal P}(S)$ with the bundle ${\cal Q}(S) \to {\cal T}(S)$ of quadratic
 differentials over Teichm\"uller space.  
 
An alternative description of ${\cal P}(S)$ was given by Thurston.
Consider a hyperbolic surface $X \in {\cal T}(S)$ (a Riemann surface) 
and a simple closed geodesic $\gamma$. Cut the surface 
at $\gamma$ and graft a flat cylinder $\gamma \times [0, t]$ of height $t$.
The result is the surface $\hbox{Gr}_{t\gamma}X$. 
In fact $\hbox{Gr}_{t\gamma} X$ has a
canonical projective structure provided by the Fuchsian uniformization of $X$
and the Euclidean structure on the cylinder.   
A simple closed geodesic
on a hyperbolic surface with a weight $t$ is the simplest example
of a measured geodesic lamination $\lambda = t\gamma$ on the surface.  
 A more general measured geodesic lamination $\lambda$ is a collection
of non-intersecting simple closed geodesics $\gamma_i$ 
each one with a weight $t_i$.  We then have
\be
\lambda = \sum t_i \gamma_i\,.
\ee
Grafting places a cylinder $\gamma_i \times [0, t_i]$ 
on each geodesic $\gamma_i$ resulting in a new surface with a projective structure.

Thurston showed that
grafting, yielding a projective structure,  can be defined more 
generally on the space ${\cal ML}(S)$ of measured geodesic laminations.  
In fact one has a map
\be
\hbox{Gr:}  \ {\cal ML}(S) \times {\cal T}(S) \to {\cal P}(S)\,.
\ee 
If we pick a surface $X\in {\cal T}(S)$ and a lamination $\lambda \in {\cal ML}(S)$ 
then $\hbox{Gr}_\lambda X$ is the projective structure we obtain, and 
we call $\hbox{gr}_\lambda X$ the conformal structure underlying the
projective structure.   

Given a Riemann surface $\Sigma$, that is, a complex structure, the surface
admits a unique hyperbolic structure.  But there are infinitely many projective
structures on $\Sigma$.  The metric that characterizes a projective structure
is the Thurston metric.   To define it, consider first a unit disk
$D= \{ z\in \mathbb{C} , |z|< 1\}$ and let $\rho_D$ denote the hyperbolic metric
on $D$.  Now, let $R$ denote a  
$\mathbb{CP}^1$ manifold, $x\in R$ denote a point, and $v\in T_x R$ a 
vector.  The Thurston metric $t_R$ assigns to the vector $v$ the length
$t_R(v)$ given by
\be
t_R(v) =:  \inf_{f : D \to R}  \rho_D(f^*v) \,,
\ee   
with the infinimum evaluated over all projective immersions $f: D \to R$
for which $x\in f(D)$.   This definition is analogous 
to the definition of the Kobayashi metric on a Riemann surface, in which case the immersions $f$ are holomorphic and the resulting Kobayashi metric coincides with the hyperbolic metric.  
The Thurston metric for the projective structure
$Gr_\lambda X$ has been shown to be the mixed metric:  hyperbolic on the vertices
and flat on the cylinders.  

Now let us examine the question of sections in the bundle 
$\chP_{g,n}$ over $\cM_{g,n}$. 
A useful result by Dumas and Wolf~\cite{dumas_wolf} (Theorem 1.1)
tells us that for any $X\in {\cal T}(S)$ and any lamination $\lambda\in {\cal ML}(S)$,
the grafting map $\hbox{gr}_\lambda X$ is a {\em homeomorphism} from ${\cal ML}(S) \to {\cal T}(S)$.   Consider now our construction of Feynman graphs with
$(t_i, \theta_i)$ parameters with {\em fixed} $\theta_i$.  We can implement all the
twists before grafting, by cutting open the geodesics and twisting.  Then we
graft with cylinders of circumference $L$ and heights 
$t_1, \cdots, t_{N_p} \in [0,\infty)$.
Since the heights are the parameters of a measured lamination, this shows that
the parameter space of heights (with fixed twists) is mapped injectively
into Teichm\"uller space.   It is not a priori guaranteed to be an injective
map into {\em moduli} space but it is locally injective and, since it approaches 
infinity in moduli space, the map is at most finite to one.  

This is encouraging, but not sufficient as we would like to have an injective
map of the  full parameter space of heights {\em and} twists into Teichm\"uller space and then into moduli space.  The case of heights and twists together
has been considered by McMullen~\cite{mcmullen}, who's views it as a
``complex earthquake"  and shows that at least in the case of a punctured torus,
  it provides a homeomorphism to Teichm\"uller space.  
 It seems likely that for large heights 
the map from the Feynman parameters to moduli space is in fact  injective: 
for long cylinders the length/twist
coordinates approach the standard plumbing coordinates used to describe degeneration.

An injective map to moduli space is not always obtained when 
grafting with heights $[0, \infty]$ and twists $[0, 2\pi]$ on arbitrary geodesics.   In our case, if we get an injective map from the {\em full} 
Feynman parameter space into moduli space, it must
be because the vertices are hyperbolic surfaces in systolic sets.  We think it is 
likely that the Feynman rules construct sections. 

\medskip  
\noindent
\underline{ A decomposition of moduli space\hskip1pt ?} 
If the Feynman rules produce a section of $\chP_{g,n}$ over $\cM_{g,n}$ the
various Feynman diagrams define a decomposition of 
$\cM_{g,n}$:\footnote{If the
Feynman rules produce no section they would still be producing a
decomposition of a singular chain in $\chP_{g,n}$ representing 
the fundamental homology of $\cM_{g,n}$.} 
\be
\cM_{g,n} = \cV_{g,n} \cup R_1 \cup \ \cdots \cup R_{3g-3+n} \,,
\ee   
where $R_k$ denotes the set of surfaces generated by a Feynman diagram
with $k$ propagators.  If we have a section, the sets on the right-hand 
side are disjoint, except at boundaries, and together built the moduli space. 
The geometric master equation implies that the
right-hand side builds a space whose only boundary is the boundary of moduli space.

We can think of $\cV_{g,n}$ as a graph with just
a vertex and $n$ external lines.  It is an $(6g-6+2n)$ real-dimensional
subset of $\cM_{g,n}$, representing the `thick' part of the moduli space.
In $R_1$ we have one propagator, representing 
grafting with parameters $(t, \theta)$ with $t\in [0, \infty), \theta \in [0, 2\pi)$. 
Points in $R_1$ arise in two possible ways.  We may have two surfaces grafted together: one in $\cV_{g_1, n_1}$ the
other in $\cV_{g_2, n_2}$ with $g_1+ g_2 = g, \, n_1 + n_2 = n + 2$, in which
case the graph is one with two vertices and one edge, as well as $n$ external
lines.  One can also have a surface in $\cV_{g-1, n+2}$ with two boundaries
grafted by the propagator, in which case the graph has one vertex, one edge
starting and ending on the vertex, and $n$ external lines.  In both cases the
assumption of a section implies that the space of
grafting parameters and vertex parameters is mapped injectively
into moduli space. 

Each  diagram in $R_{3g-3+n}$ is built with $(2g-2+ n)$ hyperbolic vertices
$\cV_{0,3}$ (having no parameters) 
and $(3g-3+n)$ propagators, assembled together 
with the instruction of a connected cubic graph.  One must then sum
over all possible inequivalent graphs.  If we have a section,
this construction maps the full parameter space of heights
and twists of each diagram injectively into the moduli space $\cM_{g,n}$
 of genus $g$ surfaces.  Each diagram builds a different region of $\cM_{g,n}$.
 The heights and twists, $(3g-3+n)$ of each, provide coordinates on the
 part of the moduli space they produce. 
 
 This is quite different from the way Fenchel-Nielsen coordinates work.
 In this case a {\em single} connected cubic graph with $(2g-2+ n)$ vertices 
 is used to glue together hyperbolic pants.  The length and twist coordinates
 are encoded on the $(3g-3+2n)$ edges of the graph, 
 each one representing a seam.  
 As these parameters run over their usual 
 ranges (zero to infinity for length and minus to plus infinity for twist) 
 one has a homeomorphism to  {\em Teichm\"uller} space.  
 One need not sum over graphs.   In the string motivated construction,
 one sums over all cubic graphs with fixed hyperbolic pants and the
 coordinates arise from grafting.

 \section{Comments and open questions}

We begin with some comments and  elaborations on our results. 

\begin{enumerate}

\item  This proposal seems to give the first rigorous 
explicit construction of string vertices.  
In~\cite{Moosavian:2017qsp,Moosavian:2017sev} some aspects of the 
construction can only be made explicit in approximations.  
The systolic minimal
area metrics, despite recent progress, have not yet been rigorously
proven to exist in the thick parts of the higher-genus moduli spaces.  

\item  A particularly elegant property of the hyperbolic vertices is
that they come naturally with collars that prevent the creation of 
short curves when vertices are glued together.  In the minimal 
area approach, one must retract the obvious local coordinates by the 
addition of stubs in order to prevent this from happening. 

\item While the construction of the quantum theory requires $L \leq L_*$, 
the classical hyperbolic closed string field theory works with arbitrary $L>0$.
That is, if we restrict the string vertices  to the genus zero contributions
$\cV = \sum_{n=3}^\infty  \,  \cV_{0,n}$, 
we get a solution of the classical master equation
\be
\label{m-eqn-cl}
\partial \cV  + \tfrac{1}{2} \{ \cV\,, \cV \} = 0 \,, 
\ee
 for {\em any} $L > 0$. 
This solution works for $L > L^*$ because, as discussed in the
proof of Theorem 1 in section~\ref{hyp-str-ver},
no non-contractible curve shorter than $L$ is created
by the gluing of two separate surfaces belonging to vertices.
While collars become tiny as $L$ grows, collars are
not needed in this case.   

In fact since  
the hyperbolic vertices (viewed as surfaces with boundaries) have finite 
constant area, as $L \to \infty$ they become like ribbon
graphs of vanishing width.  In this limit the surfaces
with infinite cylinders attached, and rescaled by $2\pi/L$, 
become the minimal area vertices
of closed string field theory.  The surface acquires
a Strebel differential whose critical graph is a restricted polyhedron
of classical closed string field theory~\cite{Kugo:1989aa,Saadi:1989tb}.

\item  Open string field theory can also be defined using hyperbolic  
metrics.  Consider the classical theory only.  The three open-string vertex
would be defined precisely by  the hyperbolic hexagon 
$\tfrac{L}{2}, \gamma , \tfrac{L}{2}, \gamma ,\tfrac{L}{2}, \gamma$,  shown in 
Figure~\ref{fig:2hex}(a) and in Figure~\ref{fig:HVert} for the case $L= L_*$.  The 
external open strings are now three semi-infinite strips of
width $L/2$ attached to the three sides of length $L/2$.  The open string
boundary conditions apply to the three $\gamma$ edges.   This 
hexagon diagram
is obtained by cutting the pants $\cV_{0,3}$ across a line invariant under an antiholomorphic involution.  In fact $\cV_{0,3}$ was defined by the gluing
of two hexagons.
The hyperbolic open string vertex does not satisfy strict associativity
and therefore hyperbolic classical 
hyperbolic open string field theory is a non-polynomial theory organized
by an $A_\infty$ algebra~~\cite{Gaberdiel:1997ia}.
It has vertices with $n\geq 4$ open strings.   
These vertices can be obtained from the closed string vertices $\cV_{0,n}$ 
by selecting the surfaces with antiholomorphic involution and cutting.

 \end{enumerate}

We end with a discussion of some open questions that seem relevant to us.
\begin{enumerate}

\item Proving that the hyperbolic string theory
 generates a section in the $\chP_{g,n}$ bundle over $\cM_{g,n}$. 
 This may require extending the theory of complex earthquakes.  
An alternative  approach would involve devising a convex minimization 
problem on a Riemann surface with punctures whose then unique 
answer is the metric built with the string field theory that uses the hyperbolic vertices.

\item  Developing the tools to compute string amplitudes in this
framework.  The first calculation would be to compute three-point functions
using the three-string vertex.  Presumably the simplest approach would 
be to develop the theory is holomorphic objects on the 3-holed sphere.
The operator formulation of the conformal field theory would then yield
the required amplitudes. 

\item The full quantum action is written in terms of the vertices that
are systolic subsets of moduli spaces with hyperbolic metrics.  Perhaps the evaluation of integrals over these sets can be performed using the associated
Teichmuller spaces, in the spirit of~\cite{mc-shane,mirz2,ellegard} and along the lines discussed
in~\cite{Moosavian:2017sev}. 

\item  Finding coordinates to describe the systolic sets that define the
string vertices.  If point one above is true, then these coordinates together
with the propagator parameters $(t_i, \theta_i)$ would give coordinates
all over moduli space.

\item  It's difficult to think about hyperbolic metrics on an $N=1$ 
super Riemann surface, but the key ideas used here
can expressed in purely group-theoretical terms.
The symmetry of the universal cover is $\hbox{SL}(2,R)$ and there is
map from the fundamental 
group to $\hbox{SL}(2,R)$.   
The length of the geodesic in a particular class in the fundamental group can be read off the corresponding conjugacy class in $\hbox{SL}(2,R)$.
There are uniformization theorems for $N=1$ super Riemann surfaces~\cite{Crane:1986uf} and the symmetry group of the universal cover is some supergroup.
 While we can't really talk about the length of a hyperbolic geodesic super Riemann surface, we can look at the conjugacy class in this super-group as a replacement.

\end{enumerate}

\subsection*{Acknowledgements} 
We are very grateful for instructive and detailed
correspondence with  J\o rgen Andersen, David Dumas, Gabriele Mondello, Michael Wolf,  and  Scott Wolpert.  

We acknowledge the hospitality of the Simons Center for Geometry and Physics
for the workshop `String field theory, BV quantization, and moduli 
spaces' (May 20-24, 2019), where this research was started.

K.C. is supported by the Krembil Foundation, the NSERC
Discovery Grant program and by the Perimeter Institute for Theoretical
Physics. Research at Perimeter Institute is supported by the
Government of Canada through Industry Canada and by the Province of
Ontario through the Ministry of Research and Innovation.
B.Z is partially supported by the U.S. Department of Energy, Office of Science, Office of High Energy Physics of U.S. Department of Energy under grant Contract Number  DE-SC0012567.

\end{document}